\documentclass[10pt,letterpaper]{article}

\usepackage{changepage}

\usepackage[utf8]{inputenc}

\usepackage{textcomp,marvosym}

\usepackage{fixltx2e}

\usepackage{amsmath,amssymb}

\usepackage{cite}

\usepackage{nameref,hyperref}

\usepackage{microtype}
\DisableLigatures[f]{encoding = *, family = * }

\usepackage{rotating}

\usepackage[aboveskip=10pt,labelfont=bf,labelsep=period,justification=raggedright,singlelinecheck=off]{caption}

\makeatletter
\renewcommand{\@biblabel}[1]{\quad#1.}
\makeatother

\date{}

\usepackage{lastpage,fancyhdr,graphicx}
\usepackage{epstopdf}
\newcommand{\etc}{etc.}
\newcommand{\etal}{~\textit{et al.}~}

\newcommand{\II}{\mathbf{I}}
\newcommand{\AAA}{\mathbf{A}}
\newcommand{\KK}{\mathbf{K}}
\newcommand{\KKi}{\KK^{-1}}
\newcommand{\HH}{\mathbf{H}}

\newcommand{\defn}[1]{\textbf{#1}}
\newcommand{\die}{\delta_i^{\mathrm{e}}}
\newcommand{\dde}{\delta^{\mathrm{e}}}

\newcommand{\ep}{\mathrm{P}} 
\newcommand{\ed}{\mathrm{D}} 
\newcommand{\er}{\mathrm{R}} 

\begin{document}
\vspace*{0.35in}

\begin{flushleft}
{\Large
\textbf\newline{The Multi-Scale Network Landscape of Collaboration}
}
\newline
\\
Arram Bae\textsuperscript{1,\Yinyang},
Doheum Park\textsuperscript{1,\Yinyang},
Yong-Yeol Ahn\textsuperscript{2},
Juyong Park\textsuperscript{1,*}
\\
\bigskip
\bf{1} Graduate School of Culture Technology, Korea Advanced Institute of Science \& Technology, Daejeon, Republic of Korea
\\
\bf{2} School of Informatics and Computing, Indiana University, Bloomington, Indiana, The United States
\\
\bigskip

%
%
\Yinyang These authors contributed equally to this work.


* juyongp@kaist.ac.kr

\end{flushleft}
\section*{Abstract}
Propelled by the increasing availability of large-scale high-quality data, advanced data modeling and analysis techniques are enabling many novel and significant scientific understanding of a wide range of complex social, natural, and technological systems. These developments also provide opportunities for studying cultural systems and phenomena -- which can be said to refer to all products of human creativity and way of life.  An important characteristic of a cultural product is that it does not exist in isolation from others, but forms an intricate web of connections on many levels.  In the creation and dissemination of cultural products and artworks in particular, collaboration and communication of ideas play an essential role, which can be captured in the heterogeneous network of the creators and practitioners of art.  In this paper we propose novel methods to analyze and uncover meaningful patterns from such a network using the network of western classical musicians constructed from a large-scale comprehensive Compact Disc recordings data.  We characterize the complex patterns in the network landscape of collaboration between musicians across multiple scales ranging from the macroscopic to the mesoscopic and microscopic that represent the diversity of cultural styles and the individuality of the artists.

\section*{Introduction}
Advances in information science and technology have enabled us to amass large-scale data from a wide range of social and cultural phenomena, stimulating the development of advanced data modeling and analysis methods for extracting useful information. This type of large-scale data collection and analysis is not limited to traditional scientific and engineering fields but is reaching into a wider range of fields such as social science and humanities, calling for deep and serious transdisciplinary effort to make full use of its universal impact~\cite{miller2008community,lazer2009computational}.  Recently, various mathematical and computational techniques have been applied to cultural data sets including recipes, music, paintings,~\etc~to gain new insights and understanding, further expanding the application in their fields~\cite{ahn2011flavor,schich2014network,bae2014network,park2014network,klingenstein2014civilizing}.

Of many new data modeling frameworks, networks in particular have gained popularity for analyzing systems whose structure and function depend critically on the connections or correlations between their components~\cite{newman2010networks,newman2003structure,faloutsos1999power,holme2004structure,jeong2000large, grimm2005pattern}. There are several essential connections between networks and culture that render such network framework necessary for a scientific understanding of culture. First, a cultural product invariably cites existing products or ideas either explicitly or implicitly. Second, most cultural products are borne out of collaborations between artists and practitioners that act as a conduit for ideas and inspirations. Accordingly there have been notable scientific studies of culture and cultural phenomena from the network perspective focusing on the relationships among cultural products, creators, consumers,~\etc~\cite{uzzi2005collaboration,salganik2006experimental,schich2012arts}.  In this paper we study the network of the creators and practitioners of culture to understand the patterns of collaborations and associations, and what they tell us about the nature of cultural prominence and diversity. Specifically we analyze the network of western classical musicians by leveraging one of the most comprehensive Compact Disc (CD) recordings databases in a rigorous fashion using established and new methods. While this paper focuses solely on music as the area of application, the analytical framework we propose should be generally applicable to any similar type of network.

Music, one of the most significant and oldest cultural traditions, boasts a rich history of cross-pollination of ideas and practices through time~\cite{struble1995history,taruskin2009music}. There have been a number of studies on musician networks: Silva~\etal studied the Brazilian popular musician network, finding basic properties such as the small-world effect and the power-law degree distribution~\cite{de2004complex}; Park~\etal considered two distinct relationship types between contemporary pop musicians (musical similarity and collaboration) and showed that they exhibit vastly different network patterns~\cite{park2007social}; Gleiser~and~Danon studied the social network of jazz musicians and found communities of musicians that correspond to regional differences and racial segregation~\cite{gleiser2003community}; and Park~\etal~studied the network of classical composers who formed communities that corresponded to a modern musicological understanding of the history of music~\cite{park2015topology}.

Those works, while having pioneered in the application of network framework to musical data, show two apparent shortcomings. First, many deal with a relatively narrow period in the history of music, mainly the latter half of the 20th century and beyond. Prominently missing is the entire body of western classical music, one of the richest musical traditions~\cite{ross2007rest,taruskin2009music}. Second, they all ignore that a musical composition is distinct from many other art forms such as paintings or sculptures in that it requires a collaboration or combination between people with differing roles, i.e. individuals or group performers, composers, conductors,~\etc~Therefore this is a heterogeneous network where the meaning of an edge depends on which node types it connects. By leveraging one of the largest databases on western classical music performance recordings that incorporates the network heterogeneity, this paper aims to shed light on the complex and heterogeneous nature of the network of collaborations in music and, more broadly, culture.  

We study the network of western classical musicians to find significant global patterns, to uncover principles that drive the connections between musicians, and to identify local network structures that allow us to represent the rich diversity within culture. Our network is constructed from ArkivMusic (http:/www.arkivmusic.com) database, an online vendor of classical music CDs.  For each CD it provides its title, release date, label, and four classes of musicians (composer, performer, conductor, and ensemble) whose compositions or performances were featured on it.  After removing the so-called compilation albums that are repackaged collections of previously released recordings, we are left with 67\,277 CDs and 75\,604 musicians, which can be represented as a bipartite network with 428\,728 edges as shown schematically in Fig. 1~(A).  Fig. 1~(B) shows a small backbone~\cite{serrano2009extracting} of the network of composers as an example (subsequent analyses are performed on the original bipartite network to minimize loss of information).  Specifically, we focus on the network patterns on three scales which we label the \emph{macroscopic}, the \emph{mesoscopic}, and the \emph{microscopic} (see Fig. 2). On the macroscopic scale we study the global, bird's-eye view of the network, which allows us to identify those musicians with universal prominence. On the mesoscopic scale we study the modular structure (node subgroups) of the network to find the strength of correlation between node characteristics and connection. Some results from the macroscopic and mesosopic scales have been previously reported by us in~\cite{bae2014network}, with some updates that reflect the most up-to-date data, although it stands on its own for a completeness and consistency leading to our new and rigorous analysis on the microscopic scale, and on the unified view of the various scales. On the microscopic scale we present how to quantify the relevance of all other musicians to a specific musician, which allows us to identify the smallest, local network landscape. Finally, we conclude by how these multiscale patterns relate to one another, letting us establish a coherent relationship between universality and diversity, two essential yet seemingly contradictory characteristics of culture.

\begin{figure}[h]
\includegraphics[width=1.0\textwidth]{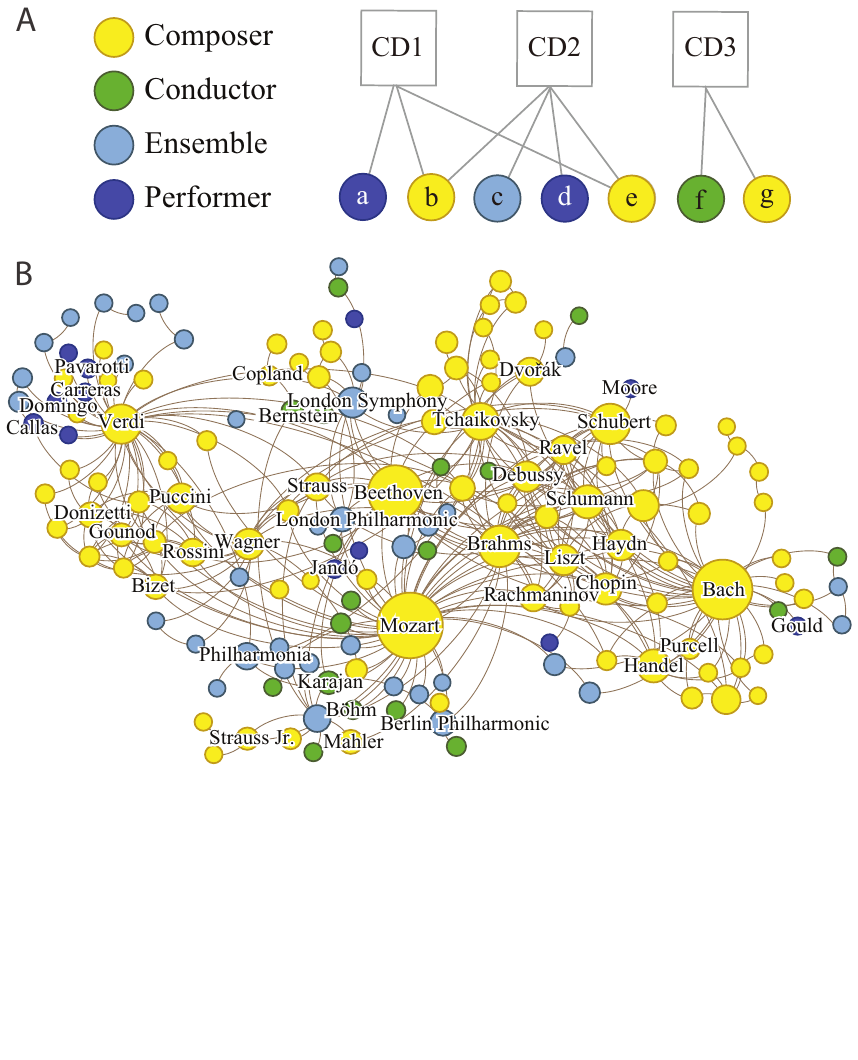}
\caption{{\bf Constructing the network of artists and cultural products.}
(A) The comprehensive classical music recordings data from ArkivMusic is a bipartite network with edges running between CDs and the musicians. The musician layer (bottom) is a heterogeneous mix of musician classes -- composers, conductors, ensembles, and individual performers. (B) A backbone of the network of musicians (CDs omitted via one-mode projection). An edge between musicians means that their compositions or performances were featured on a common CD.}
\label{00_fig}
\end{figure}

\begin{figure}[h]
\centering
\includegraphics[width=0.8\textwidth]{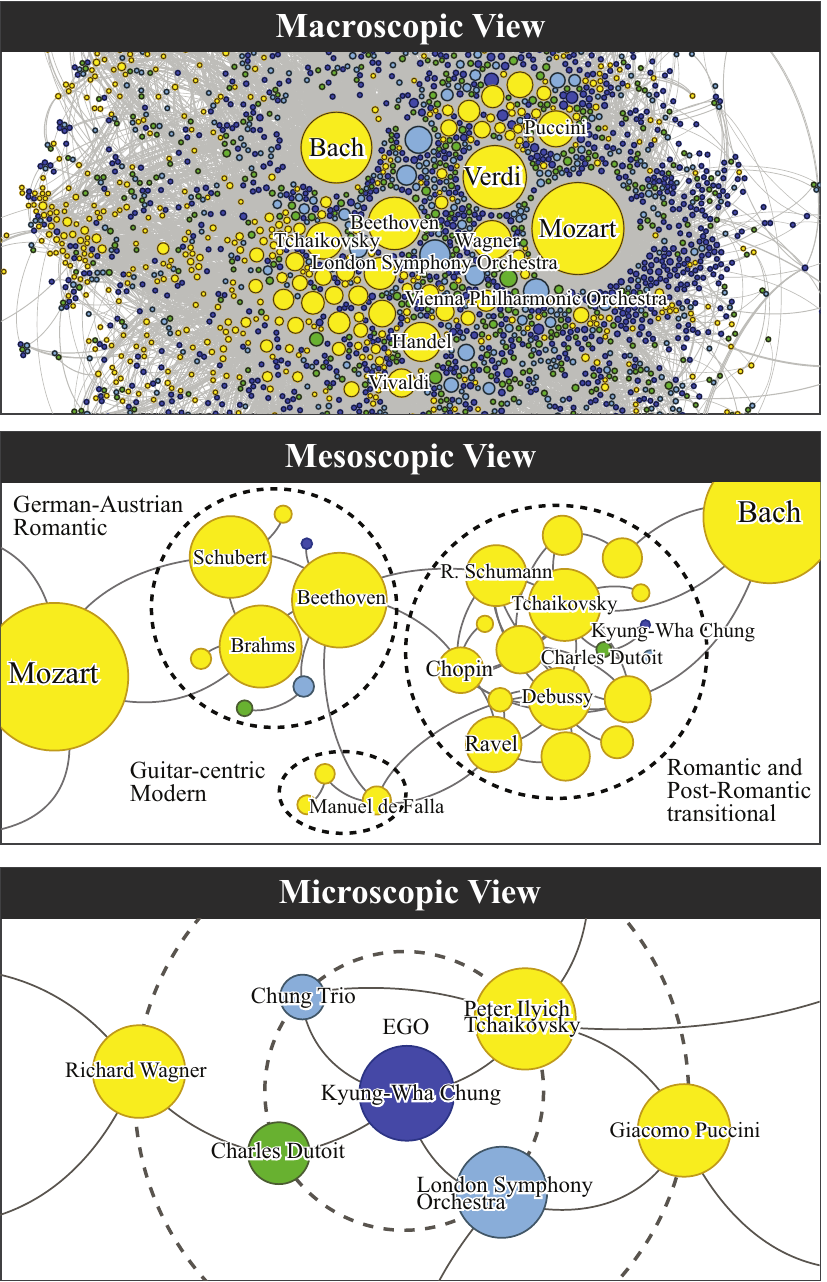}
\caption{{\bf The multiscale views of the network landscape of the classical music network.}
On the macroscopic level (top), we take a bird's-eye view of the global characteristics of the network. On the mesoscopic scale (middle), we investigate the community structure of the network that reveals the homophily based on musician characteristics such as period and nationality. On the microscopic scale (bottom), we find the local network landscape around a specific musician by quantifying the relevance of others to the musician. This type of multiscale view allows us to correctly characterize the relationships between musicians and their roles in the cultural collaboration network, where a simple global prominence (top) can easily eclipse the rich local structures that represent diverse styles (middle) and individuality of artists (bottom).}
\label{01_fig}
\end{figure}

\section*{Macroscopic Network Patterns: Global Characteristics}

On the macroscopic scale, our network shows many common characteristics of large-scale complex networks. For instance, the network possesses a giant component comprising $98.8\%$ of all nodes, meaning that most musicians are connected by a path, regardless of their active era or specialties. The average geodesic (the shortest path between two nodes) length is $5.6$ while the diameter (longest geodesic length) is $18$ in the giant component, showing the small-world property~\cite{milgram1967small,watts1998collective}. They are summarized in Table~\ref{01_tab}.

\begin{table}[!ht]
\caption{
{\bf Basic Network Properties.}}
\begin{tabular}{p{8.7cm}c}
\hline\noalign{\smallskip}
Total Number of Nodes (CDs and musicians) & 142\,881 \\
\hspace{10 pt} -- Number of CDs	& 67\,277 \\
\hspace{10 pt} -- Number of Musicians & 75\,604 \\
\hspace{20 pt} - Composers & 13\,148 \\
\hspace{20 pt} - Conductors & 5\,167 \\
\hspace{20 pt} - Performers	&	45\,907	\\
\hspace{20 pt} - Ensembles		&	11\,382	\\
Number of Edges 			&	428\,728	\\ 
Mean Degree of Musicians & 5.67	\\
Mean Geodesic Length / Diameter	&	5.6 / 18	\\
Largest Component Size &	141\,212~(98.8\%) \\
Power Exponent of Degree Distribution & $2.31\pm 0.03$ \\
\hline
\end{tabular}
\begin{flushleft}
\end{flushleft}
\label{01_tab}
\end{table}

The mean degree of musicians, i.e. the average number of CDs on which a musician's composition or performance is featured, is $5.7$. Compared with the total number of CDs $67\,277$, this tells us that the network is very sparse.  The distribution of the degree is very skewed, approximately a power law (Fig. 3~(A)). In Table~\ref{02_tab} we show, for each musician class, the ten highest-degree musicians.  For instance, Wolfgang Amadeus Mozart (1756--1791) is the most popular composer, featured on 5\,288 CDs. The tenor Pl\'acido Domingo (1941-- ) is the most popular performer, featured on 354 CDs. Herbert von Karajan (1908--1989) and London Symphony Orchestra (founded in 1904) are most prolific recording conductor and ensemble, respectively.  If we accept the degree as a simple measure for the importance of a musician, this appears to suggest that these few musicians dominate the rest in terms of musical importance.  But this may be the very reason why a macroscopic characteristic such as the degree distribution is insufficient in properly capturing the nuances in the role or importance of a musician; it would be absurd to assert, for instance, that vocalists are more important than organists simply because their degrees are larger. This problem is found again when we look at groups of musicians, shown in Fig. 3~(B)~and~(C). Fig. 3~(B) shows us that the degrees of composers from the Romantic period are disproportionately high, while Fig. 3~(C) shows that  highly skewed degree distributions are found amongst composers from the same period across all periods (see S1 Fig. for a similar figure for instruments). These findings suggest that it is necessary to examine the nature of groups of musicians that form the smaller-scale structures in the networks.

\begin{figure}[h]
\centering
\includegraphics[width=0.60\textwidth]{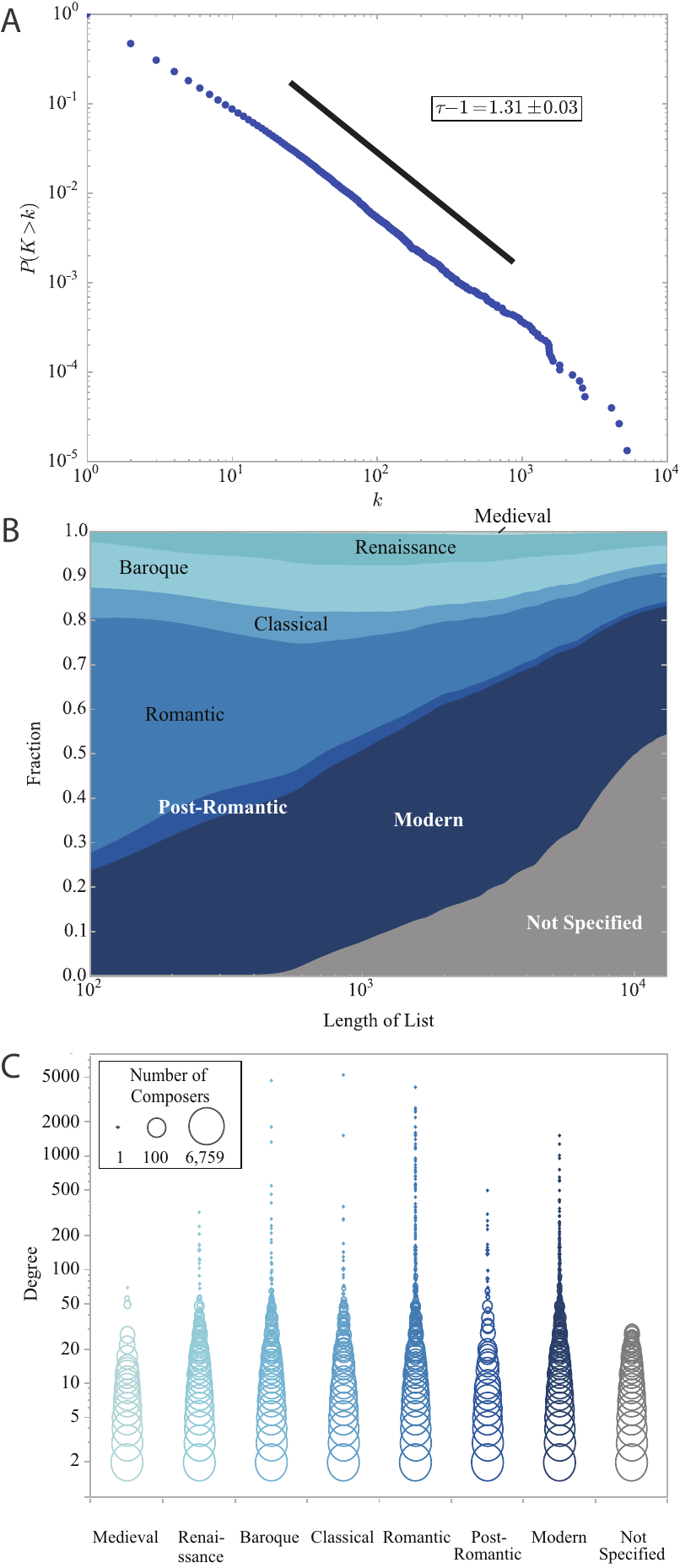}
\caption{{\bf Measuring the macroscopic network structure.}
On the macroscopic scale the network is characterized by wide variations in the visibility of the musicians, potentially masking diversity and the existence of smaller structures. (A) The cumulative degree distribution $P(K>k)$ of nodes in the bipartite network. It appears to follow a power law $P(K>k)\propto k^{-\tau+1}$ (with $\tau=2.31\pm 0.03$), suggesting an extreme level of difference in the visibility or prominence between musicians.  (B) Composers from the Romantic period are overrepresented in the lists of highest-degree composers. For instance, nearly $50\%$ of 100 highest-degree composers are from the Romantic period (far left), while it accounts for only $10\%$ of all composers (far right). (C) Significant variations in the degree of musicians are observed within the musical period as well.}
\label{02_fig}
\end{figure}

\begin{table}[!ht]
\begin{adjustwidth}{-1.25in}{0in} 
\caption{
{\bf Ten Highest-Degree Musicians (Composer, Performer, Conductor, Ensemble).}}
\begin{tabular}{p{4cm}p{4cm}p{4cm}p{4cm}}
\hline
Composer & Performer & Conductor & Ensemble \\
\hline
W.~A. Mozart\newline(Classical) & P. Domingo\newline(Tenor)	&	Herbert Von Karajan&London Symphony \newline Orchestra\\
J.~S. Bach\newline(Baroque)	&	D. Fischer-Dieskau\newline(Bass)	&	Leonard Bernstein&Vienna Philharmonic \newline Orchestra\\
L. Van Beethoven\newline(Romantic)	&	M. Callas\newline(Soprano)&Sir Neville Marriner&Philhamonia Orchestra\\
J. Brahms\newline(Romantic)	&	L. Pavarotti\newline(Tenor)&Claudio Abbado&Berlin Philharmonic \newline Orchestra\\
F. Schubert\newline(Romantic)	&	P. Schreier\newline(Tenor)&Eugene Ormandy&London Philharmonic \newline Orchestra\\
G. Verdi\newline(Romantic)	&	S. Richter\newline(Piano)&Daniel Barenboim&Royal Philharmonic \newline Orchestra\\
P.~I. Tchaikovsky\newline(Romantic)	&	J. Jando\newline(Piano)&Neeme J\"arvi&English Chamber \newline Orchestra\\
G. F. Handel\newline(Baroque)	&	N. Gedda\newline(Tenor)&James Levine&Academy of St.Martin \newline in the Fields\\
R. Schumann\newline(Romantic)	&	E. Schwarzkopf\newline(Soprano)&Sir Colin Davis&New York \newline Philharmonic\\
F. Chopin\newline(Romantic)	&	J. Sutherland\newline(Soprano)&Sir Georg Solti&Milan Teatro Alla Scala \newline Orchestra\\
\hline
\end{tabular}
\begin{flushleft}
\end{flushleft}
\label{02_tab}
\end{adjustwidth}
\end{table}

\section*{Mesoscopic Network Structures: Communities}
The previous analysis lets us see the global, system-wide characteristics of the network such as the existence of a giant component and the small-world property.  The most notable shortcoming was that a few individuals and groups appeared to be dominating the network, masking other important players in music (Fig. 3).  According to a modern understanding of networks, in fact, the small-world property by no means rules out interesting local structures in a network that represent groups of nodes called ``modules'' or ``communities.'' A common definition of a network community is a group of nodes that are more densely connected between themselves than to the rest of the network.  Communities are therefore a way of partitioning a network into meaningful mesoscopic substructures. Of many useful algorithms for identifying communities~\cite{newman2004finding,blondel2008fast,fortunato2010community,ahn2010link}, we apply the Louvain algorithm to our network~\cite{blondel2008fast} which yields 614 communities (see S2 Fig. for more information).

In Fig. 4 we show the four largest communities, along with their notable musicians' names. An examination of these suggests a positive correlation between musician characteristics and community structures, a sign of homophily or assortative mixing~\cite{newman2003social,park2007distribution,newman2002assortative,park2003origin}. For instance, community A contains many Austrian-German Romantic composers such as Ludwig van Beethoven (1770--1827), Franz Schubert (1797--1828), and Johannes Brahms (1833--1897). Community B, on the other hand, contains Aaron Copland (1900--1990), Samuel Barber (1910--1981), and John Cage (1912--1992) all prominent US-born Modern composers. To properly characterize a community in terms of such musician attributes as nationality and period, we use the following $Z$-score to quantify the degree of \emph{overrepresentation} of a musician attribute $a$ in community $s$:
\begin{align}
	Z_a^s \equiv \frac{n_a^s-np_a}{\sqrt{np_a(1-p_a)}},
\end{align}
where $n_{a}^s$ is the number of musicians with attribute $a$ inside community $s$, $n$ is the number of all musicians in the network, and $p_a$ is the fraction of musicians who have attribute $a$ in the network.  The results for three musician attributes (composer's period, performer's position, and musicians's nationality applicable to all musicians except ensembles) are shown in the boxes in Fig 4, the areas being proportional to the Z-scores.  They confirm our previous observations: Romantic composers from Austria and Germany are the most overrepresented in community A, while Modern composers from the USA are so in community B. Community C is another interesting case, with the piano being the most prominent instrument and representing the transitional period spanning late Romantic, Post-Romantic, and early European Modern, with Fr\'ed\'eric Chopin (1810--1849), Peter Ilyich Tchaikovsky (1840--1893), Claude Debussy (1862--1918), and Maurice Ravel (1875--1937).   Community D, on the other hand, represents the guitar with notable names including Issac Alb\'eniz (1860--1909, Spain), Heitor Villa-Lobos (1887--1959, Brazil) and Andr\'es Segovia (1893--1987, Spain). It also shows the dominance of Spain and Latin America, known for boasting a strong guitar tradition in modern times. Community D also clearly demonstrates the importance of local structures in understanding how diversity is represented in a cultural network: While undoubtedly a significant component of music, musicians specializing in the guitar are absent or severely underrepresented in Table~\ref{02_tab} and Fig. 3.

\begin{figure}[h]
\centering
\includegraphics[width=1.0\textwidth]{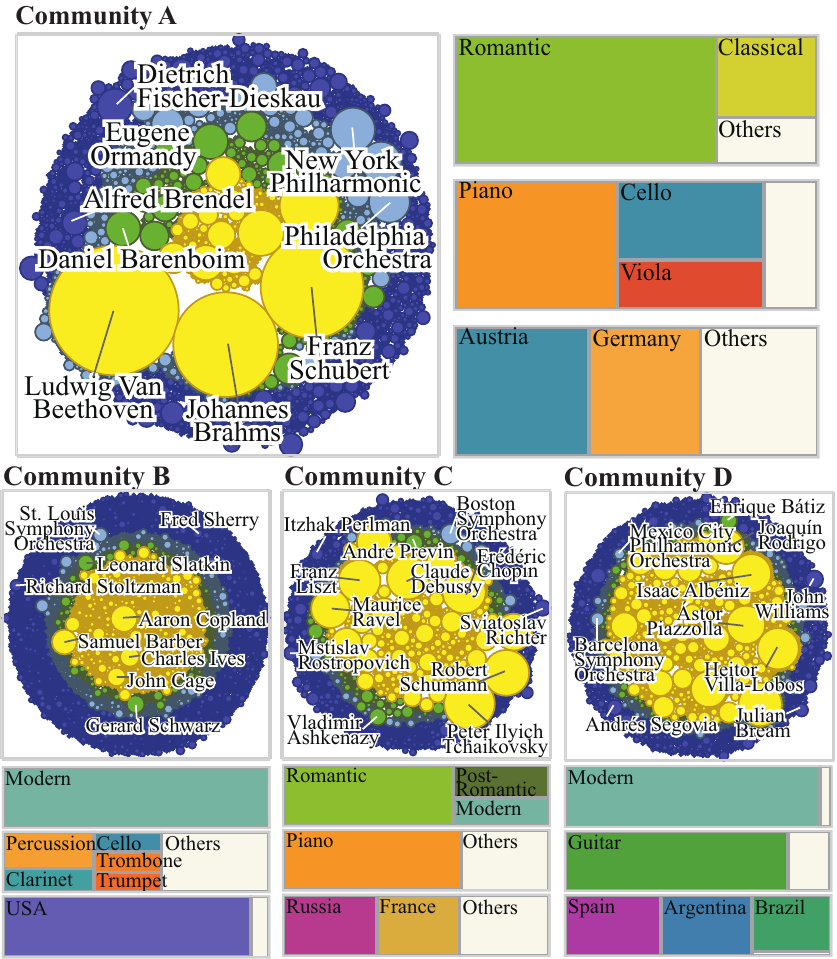}
\caption{{\bf Communities showing the mesoscopic network structure.}
On the mesoscopic scale the network is characterized by tightly-knit communities. We show four major communities. We show which musician attributes (composer periods, performer positions, and musician nationalities) are overrepresented in each community. Community A, for instance, represents the Austrian-German Romantic music; B represents the USA-based Modern music; C represents the transitional period between Romantic and Post-Romantic; finally, D represents the classical guitar.}
\label{03_fig}
\end{figure}

\section*{Microscopic Network Structures: Egocentric Relevance}
That we had to look into smaller-scale network structures to uncover important aspects of a cultural network prompts us to delve further into an even smaller scale. As mesoscopic means the network structure of groups of nodes, we take microscopic to mean the network structure centered on the individual node of the network. Traditionally, the network arranged around a specific node at the center is called the ``egocentric network'' and the central node the ``ego.'' Here we focus on determining the significance or relevance of network nodes to the specific ego, and what it can tell us about the nature of musical combinations. 

Perhaps the simplest sensible measure of the relevance of a node to another is the geodesic distance between the two.  But geodesic distance is well-known to be of limited use for the following reasons: First, since the geodesic distance is an integer and tend to be small due to the small-world property, very made nodes tend to be at the same distance from the ego. This results in a poor resolution, and not many interesting findings can be made.  Second, geodesic distance does not consider the existence of multiple paths between two nodes that could also indicate a varying level of relevance between the nodes.

Here we overcome both limitations via two straightforward modifications to the widely-used PageRank~\cite{newman2010networks,page1999pagerank} of Google that adopts the concept of random walk. We present the detailed steps for clarity. In PageRank, one assumes a random walker who visits the nodes in the network according to the following rule: At each time step, with probability $\alpha$ the walker follows a randomly chosen edge from the currently occupied node (the ``walk'' dynamic), or with probability $1-\alpha$ it jumps to a randomly chosen node in the network (the ``jump'' dynamic, no edge necessary).  After a very large number of movements, the PageRank of a node is equal to its occupation probability.

PageRank in this original form is still a global measure (its Pearson Correlation Coefficient with the degree is $0.99$ in our network), necessitating modifications to measure node's relevance to the ego. This is achieved by modifying the jump dynamic so that the walker jumps to the ego only.  This functions to reposition the walker onto the ego so that a node close to ego as well as having more paths leading to it will be visited more often, overcoming the aforementioned shortcomings of geodesic distance.  The resulting occupation probability we call \defn{Egocentric PageRank} (EP) $\ep_i^e$ defined for node $i$ and ego $e$, which can be mathematically represented as $\ep_i^e = \alpha\times\sum_j\frac{A_{ij}}{k_j}\ep_j^e+(1-\alpha)\times\die$ where $A_{ij}$ is the adjacency matrix, $k_j$ is the degree, and $\delta_i^e$ is the Kronecker delta. In vector and matrix form, it is
\begin{eqnarray}
	\vec{\ep}^e = (1-\alpha)\bigl(\II-\alpha\AAA\KKi\bigr)^{-1}\cdot\vec{\dde}.
\end{eqnarray}

The Pearson Correlation Coefficient between EP (averaged over all egos) and the degree is $0.19$, a much smaller value but it shows that the degree is still influential; it is the nature of the walk dynamics, where a high degree generally increases the chance of the node being occupied by the surfer.  With this in mind, we try the following modification to the walk dynamic: The walker now chooses a target node with a probability inversely proportional to its degree.  We call the resulting occupation probability the Degree-Neutralized Egocentric PageRank (DNEP) $\ed_i^e = \alpha\times k_i\sum_jA_{ij}H_j\ed_j+(1-\alpha)\times\die$, given in vector and matrix form as
\begin{eqnarray}
	\vec{\ed}^e=(1-\alpha)\bigl(\II-\alpha\KK\AAA\HH\bigr)^{-1}\cdot\vec{\dde}
\end{eqnarray}
where $\HH$ is a diagonal matrix of $H_j=\bigl[\sum_iA_{ij}k_i\bigr]^{-1}$, the reciprocal of the sum of the degree of node $j$'s neighbors. This degree-neutralized pairwise quantity is reminiscent of similarity measures such as SimRank proposed by Jeh and Widom~\cite{jeh2002simrank} or the regular equivalence discussed in~\cite{newman2010networks}. Note that, however, our quantity is to find the relevance (one could also say generalized closeness) between two nodes by refining PageRank, thereby not their similarity. Now the correlation between degree and DNEP of the nodes is $0.003$, showing that the degree effect has been almost eradicated. To see the difference between EP and DNEP, we define the \defn{Egocentric Relevance} (ER) $\er_i^e$ to be the linear combination of the two:
\begin{eqnarray}
	\er_i^e(\beta) = (1-\beta)\ep_i^e+\beta\ed_i^e,
\end{eqnarray}
with $\beta\in[0,1]$. ER thus changes continuously from EP to DNEP as $\beta$ is tuned from $0$ to $1$.

We now apply this method to a prominent violinist Kyung-Wha Chung (1948-- ) as an example, which is presented in Fig. 5 (see S3 Fig. for examples of other musicians).  Fig. 5~(A) shows how Chung's relevant musicians change as $\beta$ is tuned from $0$ to $1$.  When $\beta=0$ (EP), although the top ten list shares seven musicians with Table~\ref{01_tab} (Tchaikosky, London Symphony Orchestra, J.~S. Bach, Beethoven, Royal Philharmonic Orchestra, W.~A. Mozart, and Brahms), it also features those associated with her signature debut album (Andr\'e Previn and London Symphony Orchestra). EP therefore already succeeds, to some degree, in bringing forth those more intimately revelant to the ego.  The top list of relevant musicians according to DNEP is even more drastically different -- first of all, it shares no name with Table~\ref{01_tab}, and one name with (Charles Dutoit) EP (left).  On the top of the list is Chung Trio, composed of Chung and her siblings cellist Myung-Wha Chung (1944-- ) and pianist-conductor Myung-Whun Chung (1953-- ), ranked only 35th according to EP. This shows that DNEP performs even better than EP at identifying those intimately relevant to the ego: Pianist Krystian Zimerman (1956-- ) at \#4 is very well known for his Gramophone award-winning with Chung; conductor Sir Simon Rattle (1955-- ) at \#10 is famous for his work with Chung and the Vienna Philharmonic. Fig. 5~(B) shows a more extensive egocentric network landscape around Chung according to DNEP of different musician classes (with degree $30$ or larger). The distance from Chung is proportional to the log of the reciprocal of DNEP.  Among ensembles the Montreal Symphony Orchestra (founded in 1935) whose violin concerto recording with conductor Charles Dutoit (1936-- ) and Chung is very famous in the classical music community.  Among composers it is Max Bruch (1838--1920) who is the most relevant to Chung.  It is due to her recording of Max Bruch's concertos considered to be her signature achievement, pushing out better-known names in Beethoven and Bach. Chung's recordings of Jean Sibelius (1865--1957) and B\'ela Bart\'ok (1881--1945) are also famous, bringing them close to the center.

\begin{figure}[h]
\centering
\includegraphics[width=1.0\textwidth]{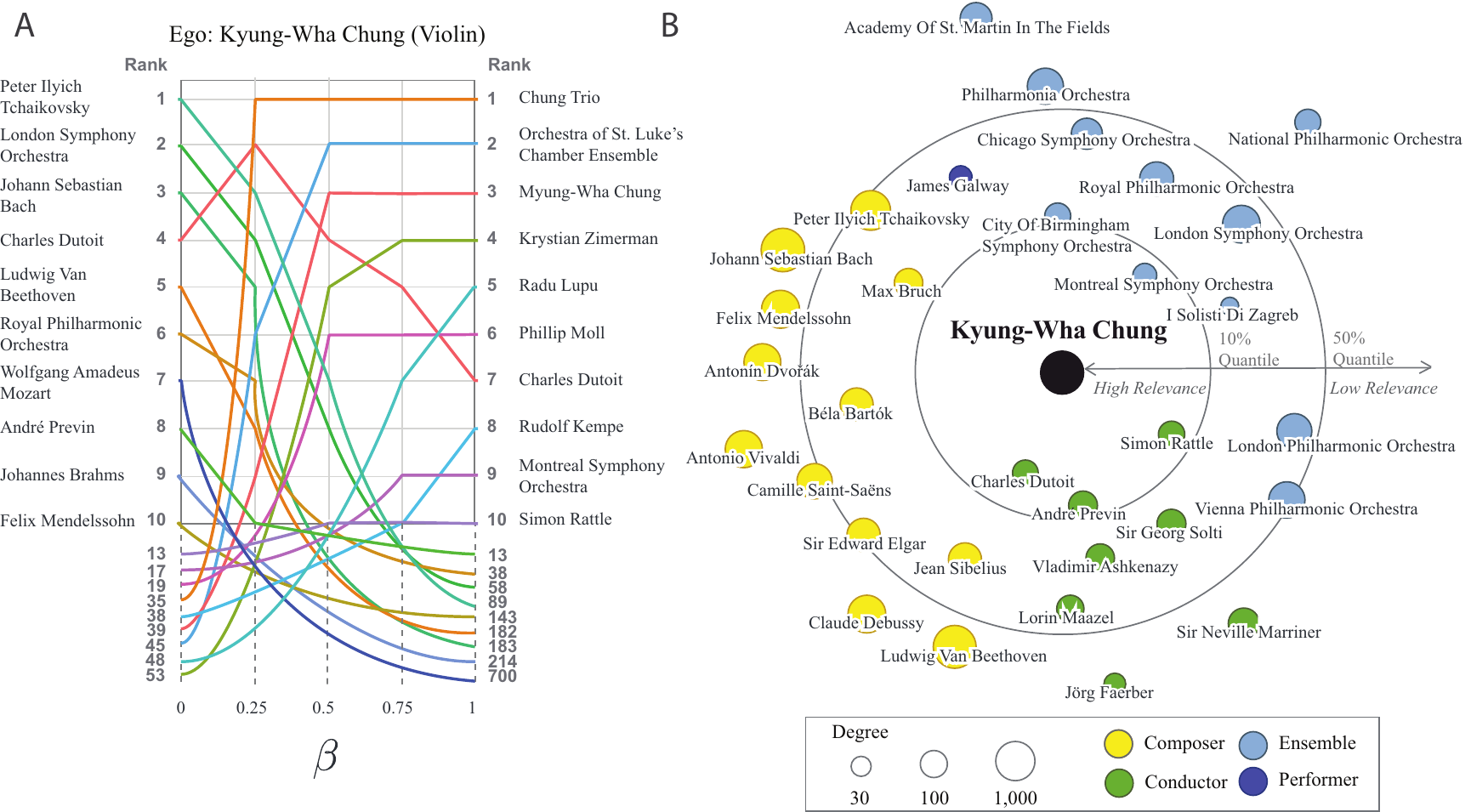}
\caption{{\bf Microscopic network structure centered on an individual musician.}
(A) The musicians most relevant to violinist Kyung-Wha Chung as an ego determined by Egocentric PageRank (EP, left) and Degree-Neutralized Egocentric PageRank (DNEP, right). Of the ten highest-EP musicians, seven (Tchaikovsky, London Symphony, J.~S. Bach, Beethoven, Royal Philharmonic, W.~A. Mozart, and Brahms) are also among the ten highest-degree nodes in the overall network.  The ten highest-DNEP musicians feature those more specific to the ego, with the Chung Trio (composed of Chung's two siblings) occupying the top spot, with Krystian Zimerman and Simon Rattle known for their collaborations with Chung in high spots. W.~A. Mozart, in contrast, falls rapidly in the ranks. (B) A figure showing the egocentric network landscape determined by DNEP $(\beta=1)$ around Kyung-Wha Chung. Highly relevant musicians tend to be lower in degree but more specifically related to her (e.g., composer Max Bruch, conductor Charles Dutoit, Montreal Symphony Orchestra,~\etc). High-degree nodes such as Tchaikovsky are pushed outwards, demonstrating the ability of DNEP to differentiate between ego-specific and universally associated musicians.}
\label{04_fig}
\end{figure}

With the success of EP and DNEP in identifying the egocentric network landscape, we now ask if we can use these measures for a group of nodes as an ego. For instance, one may be interested in those relevant to a specific instrument, not merely one individual. One possibility is to add up a musician $i$'s relevance to all nodes $e$ in the given group of interest $G$, i.e. $\sum_{e\in G}\er_i^e$.  Yet, we would also like to identify those broadly relevant to the member of $G$. We therefore propose the \defn{group-level egocentric relevance} as follow:
\begin{eqnarray}
		\er_i(G) \equiv -\biggl[\sum_{v\in G}\er_i^v\biggr]\times
	\biggl[\sum_{v\in G}\frac{\er_i^v}{\sum_{v\in G}\er_i^v}\log_{10}\frac{\er_i^v}{\sum_{v\in G}\er_i^v}\biggr],
\end{eqnarray}
a product of two terms -- the sum of relevance and an entropy-like term that gives awards those that are more uniformly relevant to the members of the group.  As an example application, we have calculated the relevance of the composers with respect to the five largest performer groups (violinists, cellists, pianists, tenors, and sopranos). Then, we took the top-100 composers in DNEP for each group, and counted how many times (one to five) they are included in the lists as a measure of the composer's versatility. The number of composers and some notable names are given in Fig. 6. We see that, perhaps surprisingly, it is only Schubert that is intimately relevant to all five groups, showing his versatility and virtuosity in both instrumental and vocal music. Mozart, Beethoven, J.~S. Bach, and Haydn are intimately relevant to four (except the cello for Mozart, and the tenor for the rest).  George Frideric Handel (1685--1759), Richard Wagner (1813--1883) and Richard Strauss (1864--1949) are intimately relevant to the tenor and the soprano, likely based on their masterpiece choral compositions and opera.  Names relevant to one group can be thought of as highly specialized composers, such as Giuseppe Tartini (1692--1770) and Niccol\`{o} Paganini (1782--1840) for the violin, and Fr\'{e}d\'{e}ric Chopin (1810--1849), Franz Liszt (1811--1886), and George Gershwin (1898--1937) for the piano.  We have to keep in mind, however, that Fig. 6 is for the five individual performer groups and that those whose major compositions were for ensembles or orchestras are likely underrepresented, which may be the case for Camille Saint-Sa\"{e}ns (1835--1921) who does not appear in the lists.

\begin{figure}[h]
\centering
\includegraphics[width=1.0\textwidth]{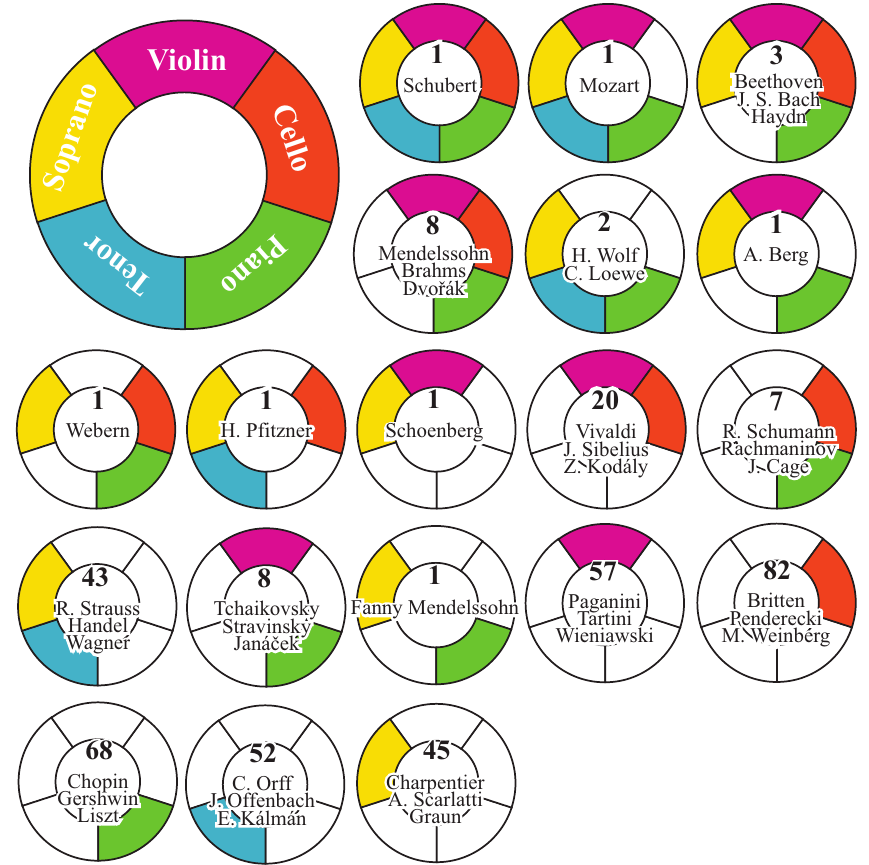}
\caption{{\bf Versatility of composers based on relevance to instrument groups.}
The number of composers highly relevant (ranked 100th or higher) to any of the five largest instrument groups (violin, cello, piano, tenor, and soprano) is in the circles. Composers relevant to multiple instrument groups in the absence of degree effect tend to be the universally recognized composers, revealing the connection between macroscopic and microscopic network patterns.}
\label{05_fig}
\end{figure}

The significance of Fig. 6 lies in the fact that it shows an explicit connection between the macroscopic and the microscopic network landscape patterns. Since the figure was based on DNEP, a measure that had nearly eradicated the degree effect, it is a representation of the local structures in the classical music network. Interestingly, however, it reproduces many names that were prominent on the macroscopic scale as those who are versatile and relevant to many classes of musicians. Fig. 6 suggests that, therefore, universality in culture stems from versatility on the microscopic level which appears as prominence on the macroscopic scale, while diversity represents the existence of many virtuosi in different subfields.

\section*{Discussion and Conclusion}
Our work shows how we can utilize the network framework to understand the landscape of cultural collaboration and combination based on large-scale databases. In order to properly understand the diversity and universality -- two of the most significant aspects of cultural creativity -- we needed to take a multi-scale view of the network, incrementally revealing the finer and more complex patterns from the network.  On the macroscopic scale we retrieve some common features of social-type network such as the power-law degree distribution and the small-world property. The inadequacy of a single-scale analysis becomes immediately clear in the beginning with the macroscopic analysis; the power-law degree distribution, for instance, suggests a strict ordering of the importance of musicians across the entire network. This is, of course, a problematic view of culture where diversity and heterogeneity are treasured.  On the mesoscopic scale we presented quantitatively the correlation between the modular structure of the network and various attribute data (periods, instruments, and nationalities), demonstrating a way to establish connection between information mined from massive digital data and a common musicological understanding of the history of music. We conducted an investigation on a further smaller scale to see how one can characterize the network properties centered on individuals. We developed two versions of egocentric relevance measures to achieve this, enabling us to discover the very musicians uniquely relevant to the ego.  This allowed us to finally understand how universality and diversity, two seemingly paradoxical nature of culture, could coexist and be represented in a coherent fasion.

We believe that our work here represents a starting point for exploring the multiscale patterns of cultural networks. With certainly a vast array of crucial questions to be explored therein, the possibility of advances in the scientific studies on cultural and humanities subjects utilizing large-scale data must be significant.


\section*{Acknowledgments}
The authors thank Joo Young Oh for helpful discussions.

\bibliography{bib/PLoSONE}

\end{document}